\documentclass[12pt,preprint]{aastex}
\usepackage{natbib}
\newcommand{\fermi}{{\it Fermi }}
\newcommand{\lat}{{\it Fermi}--LAT }
\slugcomment{Submitted to ApJ}
\shorttitle{Bright $\gamma$-ray Outbursts from 4C\,+21.35}
\shortauthors{Tanaka et al.}

\begin{document}

\title{\fermi Large Area Telescope Detection of Bright $\gamma$-ray Outbursts from a Peculiar Quasar 4C\,+21.35}

\author{
Y.~T. Tanaka\altaffilmark{1}, {\L}. Stawarz\altaffilmark{1,\,2}, D.~J. Thompson\altaffilmark{3}, F. D'Ammando\altaffilmark{4,\,5}, S.~J. Fegan\altaffilmark{6}, B. Lott\altaffilmark{7}, D.~L. Wood\altaffilmark{8}, C.~C. Cheung\altaffilmark{9}, J.~Finke\altaffilmark{8}, S. Buson\altaffilmark{10,\,11}, L. Escande\altaffilmark{7,\,12}, S. Saito\altaffilmark{1}, M. Ohno\altaffilmark{1}, T. Takahashi\altaffilmark{1}, D.~Donato\altaffilmark{13,\,14}, J.~Chiang\altaffilmark{15}, M.~Giroletti\altaffilmark{16}, F.~K.~Schinzel\altaffilmark{17}, G.~Iafrate\altaffilmark{18,\,19}, \& F.~Longo\altaffilmark{18,\,20}
}

\email{tanaka@astro.isas.jaxa.jp}

\altaffiltext{1}{Institute of Space and Astronautical Science, JAXA, 3-1-1 Yoshinodai, Chuo-ku, Sagamihara, Kanagawa 252-5210, Japan}
\altaffiltext{2}{Astronomical Observatory, Jagiellonian University, 30-244 Krak\'ow, Poland}
\altaffiltext{3}{NASA Goddard Space Flight Center, Greenbelt, MD 20771, USA}
\altaffiltext{4}{IASF Palermo, 90146 Palermo, Italy}
\altaffiltext{5}{INAF-Istituto di Astrofisica Spaziale e Fisica Cosmica, I-00133 Roma, Italy}
\altaffiltext{6}{Laboratoire Leprince-Ringuet, \'Ecole polytechnique, CNRS/IN2P3, Palaiseau, France}
\altaffiltext{7}{Universit\'e Bordeaux 1, CNRS/IN2p3, Centre d'\'Etudes Nucl\'eaires de Bordeaux Gradignan, 33175 Gradignan, France}
\altaffiltext{8}{Space Science Division, Naval Research Laboratory, Washington, DC 20375, USA}
\altaffiltext{9}{National Research Council Research Associate, National Academy of Sciences, Washington, DC 20001, resident at Naval Research Laboratory, Washington, DC 20375, USA}
\altaffiltext{10}{Istituto Nazionale di Fisica Nucleare, Sezione di Padova, I-35131 Padova, Italy}
\altaffiltext{11}{Dipartimento di Fisica ``G. Galilei", Universit\`a di Padova, I-35131 Padova, Italy}
\altaffiltext{12}{CNRS/IN2P3, Centre d'\'Etudes Nucl\'eaires Bordeaux Gradignan, UMR 5797, Gradignan, 33175, France}
\altaffiltext{13}{Center for Research and Exploration in Space Science and Technology (CRESST) and NASA Goddard Space Flight Center, Greenbelt, MD 20771, USA}
\altaffiltext{14}{Department of Physics and Department of Astronomy, University of Maryland, College Park, MD 20742, USA}
\altaffiltext{15}{W. W. Hansen Experimental Physics Laboratory, Kavli Institute for Particle Astrophysics and Cosmology, Department of Physics and SLAC National Accelerator Laboratory, Stanford University, Stanford, CA 94305, USA}
\altaffiltext{16}{INAF Istituto di Radioastronomia, 40129 Bologna, Italy}
\altaffiltext{17}{Max-Planck-Institut f\"ur Radioastronomie, Auf dem H\"ugel 69, 53121 Bonn, Germany}
\altaffiltext{18}{Istituto Nazionale di Fisica Nucleare, Sezione di Trieste, I-34127 Trieste, Italy}
\altaffiltext{19}{Osservatorio Astronomico di Trieste, Istituto Nazionale di Astrofisica, I-34143 Trieste, Italy}
\altaffiltext{20}{Dipartimento di Fisica, Universit\`a di Trieste, I-34127 Trieste, Italy}

\begin{abstract}

In this paper we report on the two-year-long \lat observation of the peculiar blazar 4C\,+21.35 (PKS 1222+216). This source was in a quiescent state from the start of science operations of the {\it Fermi Gamma-ray Space Telescope} in 2008 August until 2009 September, and then became more active, with gradually increasing flux and some moderately-bright flares. In 2010 April and June, 4C\,+21.35 underwent a very strong GeV outburst composed of several major flares characterized by rise and decay timescales of the order of a day. During the outburst, the GeV spectra of 4C\,+21.35 displayed a broken power-law form with spectral breaks observed near $1-3$\,GeV photon energies. We demonstrate that, at least during the major flares, the jet in 4C\,+21.35 carried a total kinetic luminosity comparable to the total accretion power available to feed the outflow. We also discuss the origin of the break observed in the flaring spectra of 4C\,+21.35. We show that, in principle, a model involving annihilation of the GeV photons on the He\,{\tt II} Lyman recombination continuum and line emission of ``broad line region'' clouds may account for such. However, we also discuss the additional constraint provided by the detection of 4C\,+21.35 at $0.07-0.4$\,TeV energies by the MAGIC telescope, which coincided with one of the GeV flares of the source. We argue that there are reasons to believe that the $\lesssim$\,TeV emission of 4C\,+21.35 (as well as the GeV emission of the source, if co-spatial), is not likely to be produced inside the broad line region zone of highest ionization ($\sim 10^{17}$\,cm from the nucleus), but instead originates further away from the active center, namely around the characteristic scale of the hot dusty torus surrounding the 4C\,+21.35 nucleus ($\sim 10^{19}$\,cm).

\end{abstract}

\keywords{galaxies: active --- galaxies: jets --- quasars: general --- quasars: individual(4C\,+21.35)  --- gamma rays: galaxies --- radiation mechanisms: non-thermal}

\section{Introduction}

Radio-loud active galactic nuclei (AGN) with relativistic jets oriented at small viewing angles --- called `blazars' --- are known for their particularly intense and dramatically variable $\gamma$-ray emission within the GeV photon energy range. This emission, strongly Doppler-boosted in the observer frame, dominates the total \emph{observed} radiative output of a typical luminous blazar, e.g., flat spectrum radio quasar (FSRQ), by orders of magnitude, as established by the previous studies with the EGRET instrument onboard the {\it Compton Gamma-Ray Observatory} \citep{3EGcatalog}. There is a need for careful investigation of the $\gamma$-ray properties of powerful blazar sources, enabled recently by the excellent performance of the Large Area Telescope (LAT) onboard the \fermi satellite \citep{atw09}. Still, even though hundreds of FSRQs have already been detected by LAT \citep{LATAGN}, GeV spectra of only several of the brightest ones can be studied in detail \citep{LBAS}. Moreover, until now only a few luminous (quasar-hosted) blazars have been detected at higher, TeV photon energies by the modern ground-based Cherenkov telescopes.

4C\,+21.35 \citep[also known as PKS\,1222+216, $z=0.432$;][]{1987Osterbrock} is a $\gamma$-ray-emitting FSRQ \citep{3EGcatalog,LAT1FGL}. Its peculiar large-scale ($\sim 100$\,kpc) radio structure is reminiscent of a `Wide-Angle-Tailed' morphology characteristic of radio galaxies of intermediate power located at the centers of luminous clusters, but is quite unexpected for quasar sources \citep{sai93}. At smaller (mas) scales apparent superluminal velocities have been detected for a few sub-components of the 4C\,+21.35 jet, with $\beta_{app} \gtrsim 10\,h^{-1}$ for the Hubble constant $H_{\rm 0} = h \, 100$\,km\,s$^{-1}$\,Mpc$^{-1}$ \citep{1992Hooimeyer,2001Jorstad,hom01}. The observed superluminal blobs exhibit non-radial motions, involving changes in the position angles and perpendicular acceleration \citep{hom01}. The unresolved self-absorbed radio core is characterized by a flat spectrum and a brightness temperature of $T_b \sim 5 \times 10^{11}$\,K \citep{kov05}, which again indicates a small jet inclination and relativistic beaming effects. Quite surprisingly for a blazar source, the radio core of 4C\,+21.35 is less luminous than its large-scale structure: this object is formally a `lobe-dominated quasar', since the ratio of the core--to--extended radio fluxes at GHz frequencies is of the order of unity \citep{kha04,wan04}.

4C\,+21.35 was detected by EGRET \citep[as the most likely counterpart of 3EG\,J1224+2118;][]{3EGcatalog}, and subsequently subjected to several $\gamma$-ray studies and multiwavelength modeling \citep[e.g.,][]{nandikotkur07}. During the first three months of the \fermi mission, 4C\,+21.35 was not bright enough to be included in the \lat Bright Source List \citep{LATBSL}. It was detected, however, in the First LAT Catalog \citep[1FGL;][]{LAT1FGL}, where it was associated with 1FGL\,J1224.7+2121. A $\gamma$-ray flare from this source was noted in 2009 April \citep{2009LongoATel}. It was followed in 2009 December by an even larger flare, seen by both the {\it AGILE} Gamma-ray Imaging Detector \citep{2009AGILEATel} and the {\it Fermi} LAT \citep{2009CipriniATel}. On 2010 April 24, {\it Fermi} LAT detected a particularly strong GeV outburst from the object \citep{Donato2010ATel}. The analysis performed by \citet[]{2010NeronovATel, 2010NeronovarXiv} indicated that during the flare the $\gamma$-ray emission of 4C\,+21.35 extended up to observed photon energies greater than $100$\,GeV, i.e., up to the very-high-energy (VHE) band. At that time, accompanying brightening was reported at near-infrared frequencies \citep{2010CarrascoATel}. A second huge GeV outburst was recorded by \fermi and {\it AGILE} in 2010 June \citep{Longo2010, AGILE2010}. At the same time (2010 June 17), a prominent excess around $\sim 0.07-0.4$\,TeV photon energies was noted by the MAGIC atmospheric Cherenkov telescope from the position of 4C+21.35 \citep{MAGICATel, Aleksic}, establishing this source as the third FSRQ firmly detected in the VHE range by ground-based instruments besides 3C 279 and PKS 1510$-$089 \citep[see][respectively]{TeV2,TeV1}. Followup observations by ground-based optical telescopes \citep{2010DominiciATel, 2010NesciATel} and in the X-ray/ultraviolet domain by the {\it Swift} satellite \citep{2010VerrecchiaATel_Swift} illustrate the broad multiwavelength interest in the flaring activity of this blazar. 

In this paper, we study the temporal and spectral evolution of 4C\,+21.35 in $\gamma$ rays, as observed by LAT in the first two years of the \fermi operation. The paper is organized as follows. In \S\,2, we describe \lat observations and data reduction, as well as the main observational findings. A general discussion is presented in \S\,3, while the main conclusions are summarized in \S\,4.  A standard $\Lambda$CDM cosmology with $\Omega_{\Lambda} = 0.73$, $\Omega_{\rm M} = 0.27$, and $H_{\rm 0} = 71$\,km\,s$^{-1}$\,Mpc$^{-1}$ is assumed throughout the paper.

\section{Observations and Results}

The LAT onboard \fermi is an electron-positron pair-conversion telescope sensitive to $\gamma$ rays with energies in the range from $20$\,MeV to more than $300$\,GeV. The instrument is made of a high-resolution silicon micro-strip tracker, a CsI hodoscopic electromagnetic calorimeter, and a segmented plastic scintillator detector to identify the background of charged particles \citep{atw09}. Compared with its predecessor EGRET, the LAT has a larger field of view ($\sim 2.4$\,sr), a larger effective area ($\sim 8000$\,cm$^2$ for $>1$\,GeV on-axis photons), and an improved angular resolution ($\theta_{68} \sim 0.6^{\circ}$ at $1$\,GeV for events in the front section of the tracker). The entire sky is observed every 3 hours in a survey mode.

The LAT data presented here were collected from 2008 August 4 to 2010 August 4. Only events with energies greater than $200$\,MeV were selected to minimize the systematic uncertainties. We used only ``Diffuse'' class events, which have the highest probability of being photons. We excluded the photons arriving from zenith angles $> 105^{\circ}$ to limit contamination from Earth limb $\gamma$ rays. The collected LAT data were analyzed using the standard Science Tools package (\texttt{v9r16p1}). The Instrument Response Functions (IRF) \texttt{P6\_V3\_DIFFUSE} were used. To produce the light curves and for the spectral analysis the standard tool gtlike was used. The photons were extracted from a region of interest (ROI) centered on the source and having a radius of $10^{\circ}$. The gtlike model (source model) includes the 4C\,+21.35 point source and all the point sources from the first LAT catalog \citep{LAT1FGL} that fall within $15^{\circ}$ of the source. The model also includes a background component of the Galactic diffuse emission  (\texttt{gll\_iem\_v02.fit}) and an isotropic component  (\texttt{isotropic\_iem\_v02.txt}). The isotropic component includes both the contribution from the extragalactic diffuse emission and from the residual charged-particle backgrounds. The estimated systematic uncertainty of the flux is $10\%$ at $100$\,MeV, $5\%$ at $500$\,MeV, and $20\%$ at $10$\,GeV. 

Figure\,1 shows the weekly (7-day-bin) $\gamma$-ray light curve of 4C\,+21.35 (at photon energies $E >$ 100 MeV, extrapolating from the measurements above 200 MeV using a best-fit power law) since 2008 August 4 (MJD 54682). The flux $F_{>100{\rm MeV}}$ was nearly constant until a moderate flux enhancement was seen around MJD 55100 (2009 September 26). We define this interval as a `quiescent state' (see also Table\,1). After the enhancement, the flux gradually increased with some minor flares superimposed until the source entered a particularly active phase around MJD 55310 (2010 April 24). We define the period MJD 55100--55310 as an `intermediate state', and the following one as an `active state'. Figure\,2\,(a) shows the daily light curve (again above $100$\,MeV) starting from MJD 55277 (2010 March 22). As shown, two prominent outbursts were detected during the active state around MJD 55315 (2010 April 29) and 55365 (2010 June 18). In the figure, black arrows indicate the epochs when two VHE photons with energies of 149 and 129 GeV were detected by \lat \citep{2010NeronovATel}. A red arrow depicts the date MJD 55364.9 (2010 June 17) when MAGIC detected VHE emission from 4C\,+21.35 \citep{MAGICATel}; this coincides with the second huge outburst of the source at GeV energies\footnote{We note that, in principle, the detection of the $>100$\,GeV emission from the cosmologically distant objects (especially as distant as the source discussed here with $z = 0.432$) may provide interesting or even crucial constraints on the uncertain level of the Extragalactic Background Light in the infrared--to--UV photon energy range \citep[see in this context][]{ebl-fermi}. This issue is however beyond the scope of our paper.}. Interestingly, during the first major flare the GeV flux shows ``double-horn" structure; namely, a sharp rise by a factor of $\sim 4$ within 1 day followed by a gradual 3-day decrease, then a gradual 3-day rise followed by a rapid decrease by a factor of $\sim 6$ within 1 day. The second flare shows the same pattern as the beginning of the first flare, a fast rise by a factor of $\sim 3$ within 1 day was followed by a gradual 4-day decrease. Figure\,2\,(b) exhibits the time variation of the photon index $\Gamma$ derived by fitting the data using a single power-law function. In Figures\,2\,(c) and 2\,(d), we plot the daily changes in the photon indices as a function of the $\gamma$-ray fluxes during the two flaring states for the two major outbursts. As shown, during the second flare --- the one characterized by the asymmetric profile with the decay timescale longer than the rise timescale --- a clock-wise evolution on the $\Gamma - F_{>100{\rm MeV}}$ plane can be noted, while during the first flaring epoch (composed of two sub-flares), a hint of `harder--when--brighter' behavior can be found.

We performed a detailed spectral analysis for the accumulated dataset, investigating the spectral energy distribution (SED) of 4C\,+21.35 within the LAT energy range during each activity state of the source defined in Figure\,1 and 2\,(a). The resulting spectra are shown in Figure\,3. As shown, the spectra are in general harder during the active states when compared to the quiescent state. The quiescent SED is well described by a single power-law function with photon index $\Gamma = 2.57 \pm 0.07$. This value is in agreement with that given in 1FGL for this object, and is also a typical value for the GeV photon indices of FSRQs detected by the {\it Fermi} LAT \citep{LATAGN}. During the intermediate state, a single power-law model is significantly rejected compared to a broken power-law model (the difference of the logarithm of the likelihood fits\footnote{$-2 \Delta L=-2 \log(L0/L1)$, where $L0$ and $L1$ are the maximum likelihood estimated for the null and alternative hypothesis, respectively \citep{Mattox1996}.} $-2 \Delta L =13.2$, which can be related to the $\chi^2$ distribution using Wilks' theorem), consistent with a presence of a spectral break around $E_{\rm br} \simeq 3$\,GeV. Similarly, during the first flare a broken power-law model is favored over a simple power-law ($-2 \Delta L = 35.8$). The same is true for the second flare, albeit at a lower significance level ($-2 \Delta L =12.6$). 

Table\,2 summarizes the broken power-law fits (low- and high-energy photon indices $\Gamma_{\rm LE}$ and $\Gamma_{\rm HE}$, respectively) for the discussed epochs. It is important to investigate, however, if the curvature of the flaring GeV spectrum in 4C\,+21.35 is indeed as sharp as implied by the broken power-law fit, or if it is instead much smoother, being better characterized by, e.g., the log-parabola. The corresponding `alternative' log-parabola fits are therefore given in Table\,3. It can be seen that the log parabola is also preferred over the simple power law, though the values of $\Delta L$ are not as large as in Table\,2. To investigate whether the broken power law is a better fit to the data than the log parabola we use a $\chi^2$ criterion based on the fluxes in 12 energy bands (see Figure 3). By combining the whole data set accumulated during both the intermediate and active periods (MJD 55086--55371), we found that the broken power-law model gives $\chi^2/d.o.f$ = $10.9/8$, fully compatible with the data, while the log-parabola gives $\chi^2/d.o.f$ = $27.7/9$, corresponding to a chance probability of $\sim 0.001$ (equivalent to $\sim 3.3\sigma$), and is hence disfavored by the data. We therefore conclude that the broken-power-law model is a better representation of the source spectrum during the active periods. 

The three different SEDs of 4C\,+21.35 corresponding to different periods of the active state of the source are compared in Figure\,4 with the flaring spectrum of another famous FSRQ detected by the {\it Fermi} LAT, 3C\,454.3 \citep[`the brightest flare spectrum' of 3C\,454.3 corresponding to the epoch MJD 55166--55173; see][]{Ackermann2010}. This comparison indicates a similar break energy located around $1-3$\,GeV, but a variety of low- and high-energy spectral slopes. Note, finally, that during the flares the $\gamma$-ray continuum of 4C\,+21.35 extends up to GeV photon energies with rather flat photon indices $\Gamma \simeq 2$, and this is a quite unexpected behavior for FSRQs (see the discussion below).

\section{Discussion}

\subsection{Source Energetics}

With the assumed cosmology, the redshift of 4C\,+21.35 corresponds to a luminosity distance of $d_{\rm L} = 2.4$\,Gpc. At this distance, the $\gamma$-ray spectrum of the source extending during the flares with flat photon indices $\Gamma \simeq 2$ up to the observed photon energies of $1-2$\,GeV, where it reaches monochromatic energy fluxes of the order $E\,F_{E} \simeq 10^{-9}$\,erg\,cm$^{-2}$\,s$^{-1}$, implies isotropic GeV luminosity $4 \pi d_{\rm L}^2 \times [E\,F_{E}] \simeq 7 \times 10^{47}$\,erg\,s$^{-1}$. Anticipating a bolometric correction of the order of a few, the corresponding (`flaring') $\gamma$-ray luminosity of 4C\,+21.35 would therefore be $L_{\gamma} \simeq 10^{48}$\,erg\,s$^{-1}$. The analogous value for the quiescent state should then be one to two orders of magnitude lower (see Figure\,3). Also, since the $\gamma$-ray emission of FSRQs --- generally thought to result from inverse-Comptonization of ambient photon fields within relativistic jets at sub-pc/pc distances from the central engine \citep[e.g.,][and references therein]{ghi09,sik09} --- is beamed in the observer rest frame, the total power emitted during the major flares is 
\begin{equation}
L_{\rm em} \simeq \Gamma_{\rm jet}^{-2} L_{\gamma} \simeq 10^{46} \left({\Gamma_{\rm jet} \over 10}\right)^{-2}\,{\rm erg\,s^{-1}}\, ,
\end{equation}
where $\Gamma_{\rm jet}$ is the jet bulk Lorentz factor \citep{sik97}. Note that in the broad-band modeling of blazar sources detected by EGRET, \citet{cel08} claimed $\Gamma_{\rm jet} \simeq 15$ for 4C\,+21.35. It is interesting to compare the emerging value of $L_{\rm em}$ with the total power radiated by the central engine, i.e., by the supermassive black hole/accretion disk system in this object. 

For 4C\,+21.35, \citet{fan06} give the energy flux of the H$\beta$ emission line $F_{\rm H\beta} \simeq 3.1 \times 10^{-14}$\,erg\,cm$^{-2}$\,s$^{-1}$. This corresponds to an isotropic luminosity of $L_{\rm H\beta} \simeq 2 \times 10^{43}$\,erg\,s$^{-1}$, and hence to the total luminosity of the broad-line region (BLR) of $L_{\rm BLR} \simeq 25.3 \times L_{\rm H\beta} \simeq 5 \times 10^{44}$\,erg\,s$^{-1}$ \citep{wan04,fan06}. This value is in a good agreement with (a factor of 2 lower than) the corresponding one claimed by \citet{cel97} and \citet{cel08} for the source. The luminosity of the accretion disk illuminating and ionizing the BLR clouds in 4C\,+21.35 can thus be evaluated as, roughly,
\begin{equation}
L_{\rm disk} \simeq \xi_{\rm BLR}^{-1} L_{\rm BLR} \simeq 0.5 \times 10^{46} \left({\xi_{\rm BLR} \over 0.1}\right)^{-1}\,{\rm erg\,s^{-1}}\, ,
\end{equation}
where $\xi_{\rm BLR}$ is the fraction of the disk radiation reprocessed in the BLR (which is expected to be at the level of $10\%$). This implies that, during the $\gamma$-ray flares, the radiative output of the nuclear jet in 4C\,+21.35 is of the same order of magnitude as that of the central engine, $L_{\rm em} / L_{\rm disk} \sim (\Gamma_{\rm jet} / 10)^{-2} \, (\xi_{\rm BLR} / 0.1)$. With the conservative estimates for radiative efficiencies of an AGN jet and of a standard accretion disk, the energetics of 4C\,+21.35 therefore becomes extreme, as argued below.

First, let us note that \citet{wan04} estimated the mass of the supermassive black hole in 4C\,+21.35 as $\mathcal{M}_{\rm BH} \simeq 1.5 \times 10^8 M_{\odot}$. This gives the Eddington luminosity of $L_{\rm Edd} = 4 \pi \, G \mathcal{M}_{\rm BH} \, m_p c / \sigma_{\rm T} \simeq 2 \times 10^{46}$\,erg\,s$^{-1}$ and the accretion-related luminosity in Eddington units $L_{\rm disk} / L_{\rm Edd} \simeq 0.3 \, (\xi_{\rm BLR} / 0.1)$. The implied high accretion rate indicates the accretion disk in 4C\,+21.35 to be of the standard ``optically-thick, geometrically-thin'' structure, radiating with the efficiency of $\eta_{\rm disk} \equiv L_{\rm disk} / L_{\rm acc} \simeq 0.1$, where $L_{\rm acc}$ is the total power of the accreting plasma. The analogous radiative efficiency of a nuclear AGN jet, which could be defined as $\eta_{\rm jet} \equiv L_{\rm em} / L_{\rm jet}$ for the total jet kinetic power $L_{\rm jet}$, is a poorly known parameter, though there are arguments in favor of $\eta_{\rm jet} \lesssim 0.1$ at most, since AGN jets in luminous sources are expected to transport (and to deposit) the bulk of their kinetic energies far away from their bases \citep[on kpc-Mpc scales; see, e.g.,][and references therein]{sik00}. Hence, the efficiency of the jet production in the discussed object appears to be very high, namely
\begin{equation}
{L_{\rm jet} \over L_{\rm acc}} \sim \left({\Gamma_{\rm jet} \over 10}\right)^{-2} \left({\xi_{\rm BLR} \over 0.1}\right) \, \left({\eta_{\rm disk} \over 0.1}\right) \, \left({\eta_{\rm jet} \over 0.1}\right)^{-1} \sim 1\, ,
\end{equation}
indicating that, at least during the flares, the jet in the analyzed source carries away a total power comparable to the total accretion luminosity available to feed the outflow. As a result, the two major flares detected by \lat should be considered as being saturated at the maximum level. In other words, we should not expect to detect any flares from 4C\,+21.35 more luminous than $F_{>100{\rm MeV}} \sim 10^{-5}$\,ph\,cm$^{-2}$\,s$^{-1}$.

It is interesting to note in this context that \citet{suz10} --- who modeled the broad-band spectral energy distributions of several FSRQs detected by the {\it Fermi} LAT in a quiescent state and compared the emerging model parameters with the analogous ones obtained during the flaring epochs of the same objects --- concluded that ``the difference between the low- and high-activity states in luminous blazars is due to the different total kinetic power of the jet''. This, along with our finding regarding the energetics of 4C\,+21.35 as presented above, could therefore suggest an interesting scenario for the nature of the variability in luminous blazar sources. Namely, the two findings, when considered together, would be consistent with the situation in which during the quiescent states the total kinetic power of quasar jets constitutes only a small fraction of the accreting power ($L_{\rm jet}/L_{\rm acc} \sim 0.01-0.1$), while during the major outbursts --- triggered most likely by some sort of instabilities operating within the accretion disk --- the jets carry away almost all of the available accretion-related luminosity ($L_{\rm jet} \sim L_{\rm acc}$). Since for the quasar-hosted blazars the fraction of time in which a source is in a flaring state (i.e., when the observed flux is about one order of magnitude or more higher than the average flux of a source) is about $1\%$ \citep{tav10}, such epochs characterized by the extremely efficient jet production are not representative of the whole kinetic output of the central engine, however.

Even though 4C\,+21.35 may be considered as a peculiar blazar for a number of reasons (comparable core and extended radio fluxes, or a relatively small $\mathcal{M}_{\rm BH}$, for example), we believe that the above finding/speculation should apply to most FSRQs. That is because in terms of the standard radio-loudness parameter, this object is quite representative for its class. In particular, for the $B$-band luminosity $L_B \simeq 10^{0.4 \, |M_B| +35.6}$\,erg\,s$^{-1}$\,$\sim 10^{45}$\,erg\,s$^{-1}$ following from the absolute magnitude $M_B \simeq -23.92$ as given in \citet{wan04}, and the $5$\,GHz radio luminosity $L_R^{\rm ext} \simeq L_R^{\rm core} \sim 10^{45}$\,erg\,s$^{-1}$ as given in \citet{kha04}, one can find the radio-loudness parameter of 4C\,+21.35 source $\mathcal{R} = 10^5 \times (L_R/L_B) \sim 10^3$ being typical for jetted quasars in general \citep{sik07}.

\subsection{$\gamma$-ray Emission of 4C\,+21.35}

As described in the previous section (\S\,2), the flaring spectra of 4C\,+21.35 exhibit clear breaks at observed photon energies $1-3$\,GeV. Such breaks, which seem characteristic for bright FSRQs in general, were studied most extensively in the case of the analogous blazar 3C\,454.3. In particular, during the early \lat observations of 3C\,454.3, \citet{454first} reported on the GeV flare characterized by the rise and decay timescales of $\gtrsim 3$\,days, which clearly displayed a broken power-law form of the emission continuum with the low- and high-energy photon indices $\Gamma_{\rm LE} \simeq 2.3$ and $\Gamma_{\rm HE} \simeq 3.5$, respectively, and the break photon energy $E_{\rm br} \simeq 2$\,GeV. \citet{Ackermann2010} analyzed the subsequent major outburst of 3C\,454.3, demonstrating that despite significant flux changes in $\gamma$-rays (involving multiple flares with different profiles and flux doubling timescales possibly as short as hours), the photon break energy remained stable. Those findings are therefore compatible with the ones presented in our paper.

\citet{454first} argued that the break in the GeV spectrum of 3C\,454.3 cannot result from simple cooling effects, not only because the spectral change is larger than the one following from the simplest homogeneous emission models \citep[$\Delta \Gamma \equiv \Gamma_{\rm HE} -\Gamma_{\rm LE} \simeq 1.2 > 0.5$; although see in this context][]{rey09}, but also because the location of the break --- similar for different flares and different activity levels of the source --- is not consistent with the expected one \citep[around the observed MeV photon energies; see][]{sik09}. The latter argument holds also for 4C\,+21.35, even though in this case the relatively mild spectral change observed, $\Delta \Gamma \sim 0.5$ (see Table\,2), is seemingly suggestive of the standard cooling effects. \citet{454first} showed also that the broken power-law form of the GeV continuum in 3C\,454.3 cannot result from the absorption of the emitted $\gamma$-rays via photon-photon annihilation on the lower-frequency jet emission, or on the extragalactic background light, since in both cases the corresponding optical depths for $1-3$\,GeV photons are significantly less than unity. Finally, it was argued that the pair-production processes involving the direct X-ray emission of the accretion disk and disk corona cannot play a role in this context as well, because the blazar emission zone would then have to be located too close to the black hole (a few Schwarzchild radii). Again, the same is true for 4C\,+21.35. For all those reasons, \citet{454first} advocated for the observed break in the $\gamma$-ray spectrum of 3C\,454.3 to reflect a break in the underlying energy distribution of the radiating electrons, which is ``intrinsic'' to the particle acceleration processes taking place within the emission region.

More recently, \citet{fin10} proposed an alternative explanation for the origin of the break in the spectrum of 3C\,454.3, discussing a particular model for the broad-band emission of the source. In the model, a break around the observed GeV photon energies is due to a transition from the dominant inverse-Compton scattering (in the Thomson regime) of the direct emission from the accretion disk, to the dominant inverse-Compton scattering (taking place in the Klein-Nishina regime) of the disk emission re-processed in the BLR. The change in the dominant seed photon population is crucial, since the transition between the Thomson and the Klein-Nishina regimes in the inverse-Compton scattering process involving a single target photon field cannot result in a relatively sharp break in the emission spectrum \citep[see the discussion in][and references therein]{Ackermann2010}. However, we note that in evaluating the $\gamma$-ray output of the 3C\,454.3 jet, \citet{fin10} did not take into account any line emission of BLR clouds ionized by the accretion disk, but considered only a disk continuum Thomson-scattered by the BLR clouds as a soft photon field for the inverse-Compton radiation; in reality, the line emission may dominate over the Thomson-scattered continuum by at least one order of magnitude \citep[e.g.][and references therein]{tav08}, and this can possibly affect the model results.

This line emission, on the other hand, was analyzed carefully by \citet{pou10} in their attempts to explain the GeV breaks of bright \fermi blazars. In particular, \citet{pou10} argued that the pair production on the He\,{\tt II} Lyman recombination continuum and line emission of BLR clouds may be responsible for broken power-law $\gamma$-ray spectra of FSRQs, with spectral breaks formed around photon energies of $\simeq 5$\,GeV. The model, if correct, would imply the blazar emission zone (at least in the case of quasar sources) to be located ``inside the region of highest ionization of the BLR'', i.e. at $\lesssim 0.1$\,pc distances from the central engine. One should mention at this point that a possible challenge to the proposed scenario could be a lack of any observational signatures for the emission of secondary pairs created via the photon-photon annihilation in the spectra of FSRQs \citep[e.g., in the X-ray domain; see in this context the discussion in][]{ghi09}, but this issue would have to be investigated in a more quantitative way in the particular context discussed here before drawing any definitive statements. In addition, in the model one should also expect the Klein-Nishina effects to modify the energy distribution of the highest-energy electrons cooling radiatively via inverse-Compton upscattering of the UV photons from the BLR, and thus producing the observable signatures in the synchrotron continua of luminous blazars \citep{mod05}. Before now, no such signatures have been found \citep{sik09}, although the frequency regime of interest (UV/soft X-ray) is particularly challenging observationally. Keeping the above caveats in mind, below we discuss in more detail whether the model by \citet{pou10} can be applied to 4C\,+21.35.

First, let us therefore estimate the luminosity of the He\,{\tt II} Lyman complex in 4C\,+21.35. For a wide range of the values characterizing the ionization parameter and column density in the closest vicinities of black hole/accretion disk systems of quasar sources, one expects $L_{\rm He{\tt II}\,\,Ly}$ to be of the order of $10\%$ of the Ly\,$\alpha$ emission \citep{tav08,pou10}. Hence, for the Ly$\alpha$ luminosity of the BLR in 4C\,+21.35, namely $L_{\rm Ly\alpha} \simeq 4.5 \times L_{\rm H\beta} \simeq 10^{44}$\,erg\,s$^{-1}$ \citep[see][]{wan04}, we get $L_{\rm He{\tt II}\,\,Ly} \simeq 10^{43}$\,erg\,s$^{-1}$. Second, the characteristic scale for the BLR in the object, $R_{\rm BLR}$, should be evaluated. This can be done by applying the standard relation between the disk luminosity $L_{\rm disk}$ and $R_{\rm BLR}$ \citep[e.g.,][and references therein]{ghi08}, to obtain
\begin{equation}
R_{\rm BLR} \simeq 0.1 \left({L_{\rm disk} \over 10^{46}\,{\rm erg\,s^{-1}}}\right)^{1/2}\,{\rm pc} \simeq 2 \times 10^{17} \, \left({\xi_{\rm BLR} \over 0.1}\right)^{-1/2}\,{\rm cm}
\end{equation}
(see Eqn.\,2). Hereafter we take the characteristic size for the high-ionization lines (He\,{\tt II} Lyman complex) to be half this value, namely $\tilde{R}_{\rm BLR} \simeq R_{\rm BLR}/2$ \citep{pou10}, and fix $\xi_{\rm BLR} = 0.1$. The opacity for the few-GeV photons is then $\tau_{\gamma \gamma}(5\,{\rm GeV}) = \tau_{\rm T}^{5\,{\rm GeV}} \times (\sigma_{\gamma \gamma} / \sigma_{\rm T})$, where $\sigma_{\gamma \gamma}/\sigma_{\rm T}$ is the photon-photon pair production cross section in the units of the Thomson cross section, and
\begin{equation}
\tau_{\rm T}^{5\,{\rm GeV}} = {\sigma_{\rm T} \, L_{\rm He{\tt II}\,\,Ly} \over 4 \pi \, \tilde{R}_{\rm BLR} \, c \, E_{\rm He{\tt II}}} \simeq 2
\end{equation}
for all the parameters as discussed above and the characteristic photon energy of the He\,{\tt II} Lyman emission $E_{\rm He{\tt II}} \simeq 50$\,eV. Even though implying formally $\tau_{\gamma \gamma} \lesssim 1$ for $E \sim 5$\,GeV $\gamma$-ray photons\footnote{Note that, assuming an isotropic distribution of soft photons, the maximum of the pair production cross section $\sigma_{\gamma \gamma} \sim \sigma_{\rm T}/3$ is reached for $E_{\gamma} \, E_0 \simeq 2 m_e^2 c^4$ \citep{zdz88}.}, this condition is exactly what is needed to produce a GeV break observed in the spectrum of 4C\,+21.35, because in the analyzed model one expects $\Delta \Gamma \sim \tau_{\rm T}^{5\,{\rm GeV}} / 4 \simeq 0.5$. In other words, the fact that 4C\,+21.35 is characterized by a lower accretion disk luminosity than 3C\,454.3 is consistent --- in the framework of the discussed model --- with the fact that the spectral break in the former source is smaller than in the latter one ($\Delta \Gamma \sim 0.5$ versus $1.2$; note here that $L_{\rm Ly \alpha} \propto L_{\rm disk}$ and $R_{\rm BLR} \propto L_{\rm disk}^{1/2}$, so that $\tau_{\rm T} \propto L_{\rm disk}^{1/2}$).

We conclude that as long as the \lat observations of 4C\,+21.35 are considered \emph{solely}, the model proposed by \citet{pou10} can nicely account for the break observed in the spectrum of this blazar. An additional observational constraint is provided by the detection of 4C\,+21.35 at $0.07-0.4$\,TeV energies by the MAGIC telescope \citep{MAGICATel, Aleksic}, which coincided with the second GeV flare of the source. This coincidence suggests co-spatiality of the GeV and TeV emitting regions. Yet, if the GeV break is indeed due to the pair production on the high-ionization lines emitted by the BLR clouds, the emission zone has to be located at distances $r \simeq R_{\rm BLR} \simeq 10^{17}$\,cm from the central engine. At such distances, on the other hand, the photon-photon annihilation related to the near-infrared $\lambda_0 \simeq E_{\gamma} h/ 2 m_e^2 c^3 \sim 0.2 - 1$\,$\mu$m emission of the circumnuclear matter is likely to prevent the $E_{\gamma} \sim 0.07 - 0.4$\,TeV photons from escaping the circumnuclear environment, as argued below. 

First, let us note that the multi-body emission continuum of the standard accretion disk, peaking around $\sim 10^{15}$\,Hz frequencies, extrapolates down to near-infrared (NIR) frequencies with the $\sim 1$\,$\mu$m luminosity constituting at least $10\%$ of the total disk luminosity \cite[see, e.g.,][]{tav08}. This, together with $L_{\rm disk}$ estimated above, would then imply the NIR luminosity of the accreting matter in 4C\,+21.35 of the order of $\gtrsim 10^{45}$\,erg\,s$^{-1}$, which is in good agreement with the value $L_{\rm 2.2\mu m} \simeq 3.6 \times 10^{45}$\,erg\,s$^{-1}$ following from the $2.2$\,$\mu$m flux $F_{\rm 2.2\mu m} \simeq 3.873$\,mJy provided by \citet{fan06} for the object. Taking into account a possible contribution of the hot dust (located at some further distance, see below) and of the synchrotron jet emission to the observed NIR flux, we consider $L_{\rm NIR} \simeq 10^{45}$\,erg\,s$^{-1}$ to be a safe guess regarding the disk luminosity in 4C\,+21.35 indeed. Yet this emission is produced in the innermost parts of the accretion flow, with the characteristic size much smaller than $R_{\rm BLR}$. This implies that the NIR photons emitted directly by the disk illuminate the jet region located around $R_{\rm BLR}$ exactly from behind, what reduces significantly the efficiency of the pair production process due to the very kinematics of the photon-photon interaction \citep[see the discussion in][]{rei07}\footnote{It may be easily demonstrated that for the jet regions placed much closer to the active center than $R_{\rm BLR}$, this efficiency is not going to be drastically reduced. Therefore, a $\lesssim$\,TeV emission zone in 4C\,+21.35 located at $r \ll R_{\rm BLR}$ may be excluded.}. Hence, the opacity for the $\lesssim$\,TeV radiation produced within the BLR is dominated not by the direct NIR emission of the accretion disk, but instead by the soft infrared photons emitted by the BLR itself, and also by the circumnuclear dust distributed beyond the BLR clouds.

Regarding the former component, we refer to the detailed calculations presented in \citet{rei07} and \citet{tav09}, which are based on the working assumption about a spherical distribution of BLR clouds around the active center. These calculations show in particular that the corresponding optical depth for the $\lesssim 1$\,TeV photons is of the order a few, assuming the standard value $\xi_{\rm BLR} \sim 0.1$. This already suggests that the sub-TeV emission of 4C\,+21.35 (as well as the GeV emission of the source, if co-spatial), is not likely to be produced inside the region of the highest ionization of BLR, but instead at $r > \tilde{R}_{\rm BLR}$. Even more problematic may be however the latter component, which is related to the circumnuclear hot dust region (HDR) forming a torus-like structure around the accreting black hole. The characteristic scale of this structure, $R_{\rm HDR}$, can be estimated by applying the established relation between the disk luminosity $L_{\rm disk}$ and $R_{\rm HDR}$ \citep[hereafter we assume the hot dust temperature $T_{\rm HDR} \simeq 10^3$\,K; see][and references therein]{sik09}. For the particular parameters of 4C\,+21.35 one obtains in this way, roughly,
\begin{equation}
R_{\rm HDR} \simeq 4 \left({L_{\rm disk} \over 10^{46}\,{\rm erg\,s^{-1}}}\right)^{1/2}\,{\rm pc} \simeq 10^{19}\,{\rm cm}\,.
\end{equation}
The corresponding optical depth for the TeV photons produced within $r < R_{\rm HDR}$ is then
\begin{equation}
\tau_{\gamma \gamma}(1\,{\rm TeV}) \simeq {\sigma_{\rm T} \, \xi_{\rm HDR} \, L_{\rm disk} \over 12 \pi \, R_{\rm HDR} \, c \, E_{\rm HDR}} \sim 100\,\left({\xi_{\rm HDR} \over 0.1}\right)\, ,
\end{equation}
where $E_{\rm HDR} \simeq 0.3$\,eV, and $\xi_{\rm HDR}$ is the fraction of the disk radiation reprocessed in the inner regions of the dusty torus, expected again to be at the level of $10\%$. 

The obtained value of $\tau_{\gamma \gamma}(1\,{\rm TeV})$ is confusingly high. It may suggest that --- \emph{if} the $\gamma$-ray continuum of 4C\,+21.35 in the source rest frame extends up to $\simeq 1$\,TeV energies, while at the same time the emission zone is located at distances $r < R_{\rm HDR}$ --- the $\xi_{\rm HDR}$ parameter is significantly lower than expected (and, in fact, than inferred for the other analogous sources). Alternatively, the distribution of the hot dust around the nucleus of the discussed object may be very different from the simple geometry anticipated here (and in the literature). Nevertheless, even if both effects are relevant, the basic estimates presented above imply that one really should expect non-negligible, or even significant pair-production signatures around TeV photon energies in the intrinsic spectrum of 4C\,+21.35. Lack of such signatures would therefore place the blazar emission zone in the discussed object at relatively large distances from the active center, $r \gtrsim R_{\rm HDR}$, as recently advocated by \citet{sik09} for luminous blazars in general. On the other hand, the fact that the MAGIC spectrum of 4C\,+21.35 extends up to $0.4$\,TeV energies only, may be providing a crucial piece of evidence in this context, since the emission of the hot dust is expected to fall sharply at wavelengths $\lambda < 1$\,$\mu$m, thus decreasing the optical depth for $E_{\gamma} < 0.4$\,TeV photons.

\section{Conclusion}

In this paper we report on the two-year-long \lat observation of the peculiar blazar 4C\,+21.35 (PKS 1222+216). This source was in a quiescent state from at least August 2008 until September 2009, and then entered a period of enhanced activity consisting of a gradual flux increase with some moderate flares superimposed. In April and June 2010, 4C\,+21.35 underwent a very strong GeV outburst composed of several major flares characterized by the rise and decay timescales of the order of a day, and by the daily $\gamma$-ray fluxes nearly comparable to that of the Vela pulsar, the brightest persistent source in the $\gamma$-ray sky. During the outburst, the GeV spectra of 4C\,+21.35 displayed a broken power-law form with spectral breaks located near $1-3$\,GeV photon energies, similar to the flaring spectra of another luminous blazar, 3C\,454.3. We demonstrate that, at least during the major flares, the jet in 4C\,+21.35 carries a total power comparable to the total accretion luminosity available to feed the outflow. As a result, the two major flares detected by the {\it Fermi} LAT should be considered as being saturated at the maximum level, and we should not expect to detect any flares from 4C\,+21.35 more luminous than $F_{>100{\rm MeV}} \sim 10^{-5}$\,ph\,cm$^{-2}$\,s$^{-1}$. We also discuss the origin of the break observed in the flaring spectra of 4C\,+21.35. We show that, in principle, the model involving annihilation of the GeV photons on the He\,{\tt II} Lyman recombination continuum and line emission of BLR clouds \citep[as proposed by][]{pou10} may account for such breaks. However, we also discuss the additional constraint provided by the surprising detection of 4C\,+21.35 at $\lesssim 1$\,TeV energies by the MAGIC telescope, which coincided with the second GeV flare of the source. We argue that this sub-TeV emission (as well as the GeV emission of the source, if co-spatial), is not likely to be produced inside the region of highest ionization of BLR ($\sim 10^{17}$\,cm), because of the expected large opacity for $\gamma$-rays related to photon-photon annihilation on the infrared photon field provided by the accretion disk, BLR, and nuclear dust. Instead, it seems to originate further away from the active center, most likely as far as the characteristic scale of the hot dusty torus surrounding the 4C\,+21.35 nucleus ($\sim 10^{19}$\,cm). 

After completing this work, we learned that our basic estimates regarding the TeV opacity are supported by the most recent analysis of the {\it Spitzer Space Telescope} data by \citeauthor{malmrose} (2011), who found a prominent infrared excess in the observed spectrum of 4C\,+21.35, and showed that the excess is well modeled by the $\sim 10^3$\,K dust emission. The extremely high isotropic luminosity of the hot dust estimated by \citeauthor{malmrose} indicates that the fraction of the disk radiation reprocessed in the inner regions of the dusty torus in 4C\,+21.35 is as large as $\xi_{\rm HDR} \sim 1$. The exact spectral shape and the luminosity of the infrared continuum of the source available in a near future, as well as of the detail of the TeV spectrum soon reported by the MAGIC Collaboration (Aleksic et al., 2011, to be submitted), will allow us to re-examine the conclusions and estimates presented in our paper.

\acknowledgments

The \textit{Fermi}-LAT Collaboration acknowledges generous ongoing support
from a number of agencies and institutes that have supported both the
development and the operation of the LAT as well as scientific data analysis.
These include the National Aeronautics and Space Administration and the
Department of Energy in the United States, the Commissariat \`a l'Energie Atomique
and the Centre National de la Recherche Scientifique / Institut National de Physique
Nucl\'eaire et de Physique des Particules in France, the Agenzia Spaziale Italiana
and the Istituto Nazionale di Fisica Nucleare in Italy, the Ministry of Education,
Culture, Sports, Science and Technology (MEXT), High Energy Accelerator Research
Organization (KEK) and Japan Aerospace Exploration Agency (JAXA) in Japan, and
the K.~A.~Wallenberg Foundation, the Swedish Research Council and the
Swedish National Space Board in Sweden.

Additional support for science analysis during the operations phase is gratefully
acknowledged from the Istituto Nazionale di Astrofisica in Italy and the Centre 
National d'\'Etudes Spatiales in France.

Y.~T.~T. is supported by JSPS research fellowships for Young Scientists. \L.~S. is grateful for the support from the Polish MNiSW through the grant N-N203-380336. Y.~T.~T. and \L.~S. acknowledge Marek Sikora for his useful comments and remarks. The authors also thank A. Marscher for providing the results of the Spitzer observations of 4C\,+21.35.

\clearpage

\begin{table}

\begin{center}

\begin{tabular}{lcc}

\hline

\hline

 State & Start date (MJD) & End date (MJD)\\

 \hline

Quiescent  & 2008 August 4 (54682.66) & 2009 September 12 (55086.00)\\

Intermediate+Active & 2009 September 12 (55086.00) & 2010 June 24 (55371.00) \\

 Intermediate &2009 September 12 (55086.00) & 2010 April 23 (55309.00) \\

 First flare & 2010 April 23 (55309.00) & 2010 May 2 (55318.00)\\

 Inter-flare & 2010 May 2 (55318.00) & 2010 June 16 (55363.00) \\

 Second flare & 2010 June 16 (55363.00) & 2010 June 24 (55371.00) \\

\hline

\end{tabular}

\caption{Time intervals defined in this paper; see Figure\,1 and 2\,(a).}

\end{center}

\end{table}

\begin{table}

\begin{center}

\begin{tabular}{lcccc}

\hline

\hline

State  & $\Gamma_{\rm LE}$  & $\Gamma_{\rm HE}$ & $E_{\rm br}$      &    $\Delta L$ \\

              &                                        &                                       & [GeV]                         & \\

\hline

Intermediate+Active & 2.18$\pm$0.02 &2.64$\pm$0.06 & 2.4$^{+0.2}_{-0.2}$ & -27.2\\

\hline

Intermediate & 2.30$\pm$0.05 &2.69$\pm$0.40 & 2.5$^{+1.1}_{-0.4}$ & -6.6\\

First flare & 1.80$\pm$0.06 &2.40$\pm$0.07 & 1.1$^{+0.3}_{-0.2}$ & -17.9\\

Inter-flare & 2.24$\pm$0.03 &2.81$\pm$0.14 & 2.7$^{+0.6}_{-0.6}$    & -8.4\\

Second flare & 2.00$\pm$0.05 &2.44$\pm$0.10 & 1.7$^{+1.1}_{-0.4}$   & -6.3\\

\hline

\end{tabular}

\caption{Spectral fits using broken power-law model with low- and high-energy spectral indices $\Gamma_{\rm LE}$ and $\Gamma_{\rm HE}$, respectively, and the break photon energy $E_{\rm br}$. In the last column, $\Delta L$ is the difference of the logarithm of the likelihood of the fit with respect to a single power-law fit.}

\end{center}

\end{table}

\begin{table}

\begin{center}

\begin{tabular}{lccc}

\hline

\hline

State & $\alpha$  & $\beta$ &    $\Delta L$ \\

           &              &                    & \\

\hline

Intermediate+Active & 2.11$\pm$0.03 & 0.06$\pm$0.01 & -19.6\\

\hline

Intermediate & 2.26$\pm$0.05 & 0.04$\pm$0.02 & -3.1\\

First flare & 1.70$\pm$0.09 & 0.13$\pm$0.02       & -17.9\\

Inter-flare & 2.14$\pm$0.06 & 0.07$\pm$0.02      & -7.0\\

Second flare & 1.94$\pm$0.08 & 0.07$\pm$0.02 & -4.8\\


\hline

\end{tabular}

\caption{Alternative spectral fits using log-parabola function $dN/dE \propto \left(E/E_{\rm cr} \right)^{-\alpha-\beta \log(E/ E_{\rm cr})}$, where $E_{\rm cr}$ is fixed to 300 MeV. In the last column, $\Delta L$ is the difference of the logarithm of the likelihood of the fit with respect to a single power-law fit.}

\end{center}

\end{table}

\clearpage

\begin{figure}

\epsscale{.80}

\plotone{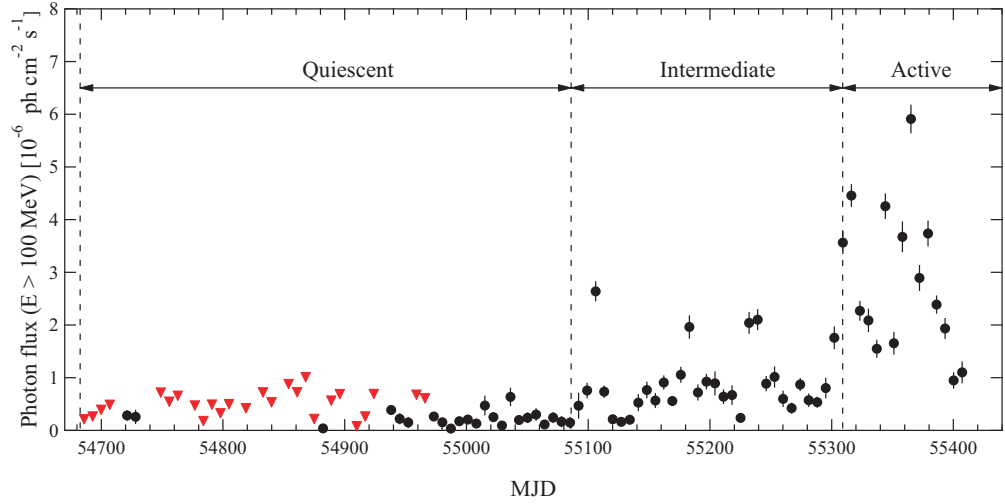}

\caption{Weekly (7-day-bin) $\gamma$-ray light curve of 4C\,+21.35 (photon energies $E > 100$\,MeV) from 2008 August to 2010 August obtained by the {\it Fermi} LAT. The $95\%$ upper limits are shown by red triangles. Three main epochs marking different activity levels of the source are defined (see also Table\,1).}

\end{figure}

\begin{figure}

\plotone{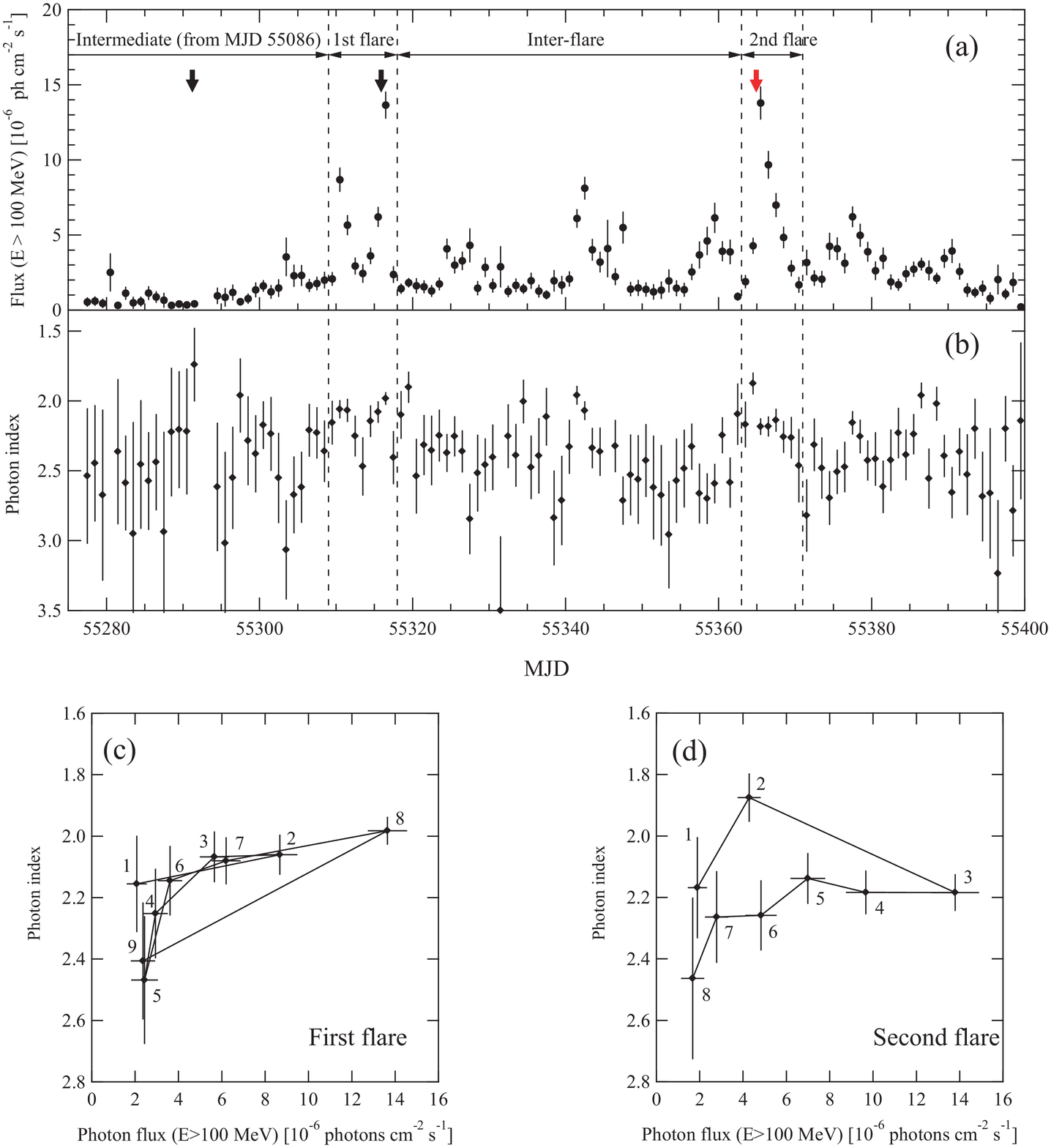}

\caption{{\bf (a)} Daily $\gamma$-ray light curve  ($E > 100$\,MeV) during an active phase of 4C\,+21.35 as observed by the {\it Fermi} LAT. Black arrows indicate the epochs when two very high energy photons with energies above $100$\,GeV were detected by the {\it Fermi} LAT \citep{2010NeronovATel}. A red arrow depicts the date MJD 55364.9 when MAGIC detected very high energy emission from 4C\,+21.35 \citep{MAGICATel}. {\rm (b)} Daily variation of the photon index derived from a single power-law fitting to the LAT data. {\bf (c)} Relation between daily $\gamma$-ray fluxes and photon indices measured during the first flare. The numbers indicate the days from the onset of the flare. {\bf (d)} The same as (c) but for the second flare.}

\end{figure}

\begin{figure}

\epsscale{0.8}

\plotone{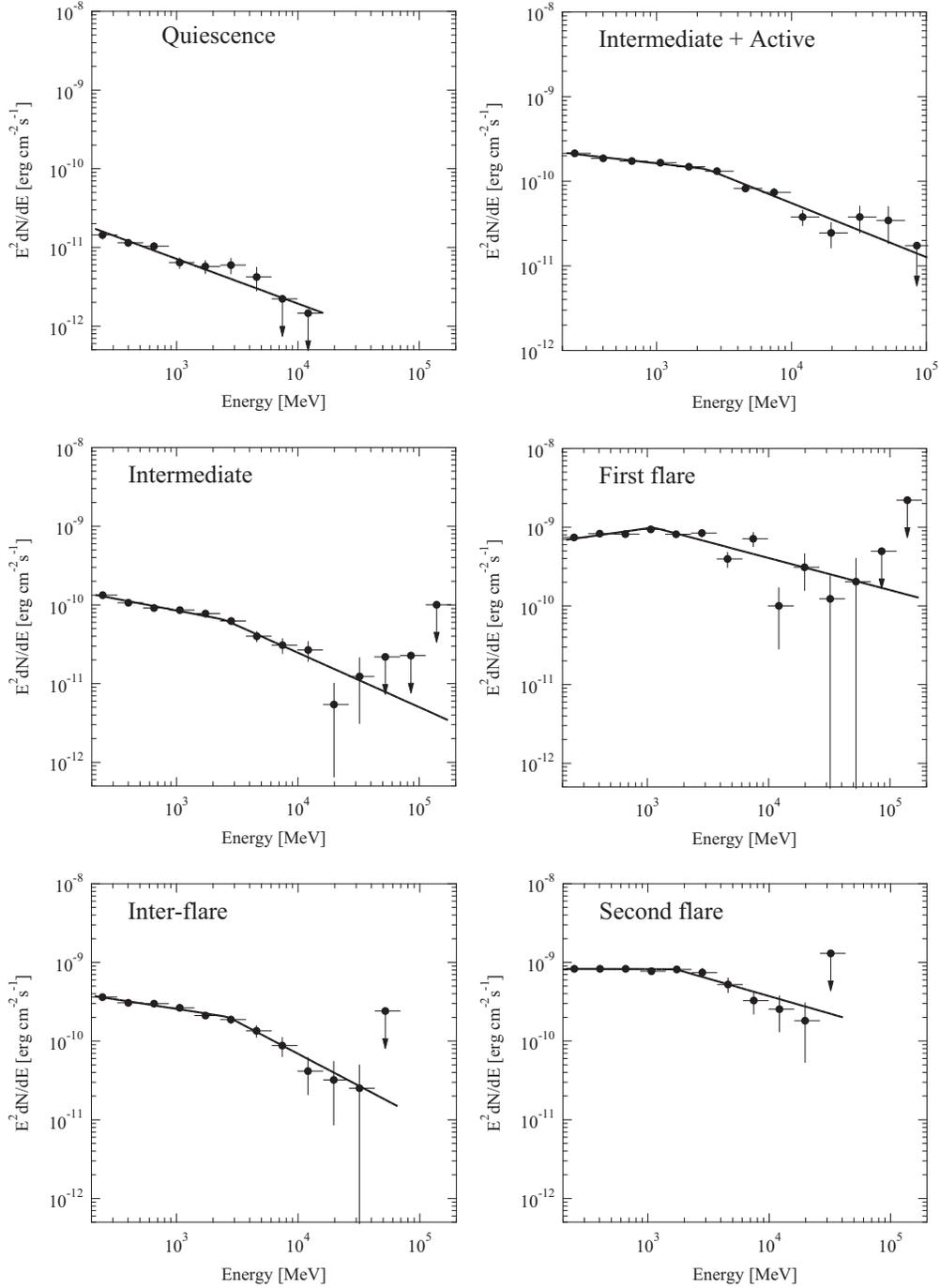}

\caption{The spectral energy distribution of 4C\,+21.35 within the LAT energy range during each activity state of the source defined in Figure\,1 and 2\,(a) (see also Table\,1). Solid lines show the best-fit broken power-law models. Arrows indicate $95\%$ upper limits for the binned spectra.}

\end{figure}

\begin{figure}

\epsscale{0.8}

\plotone{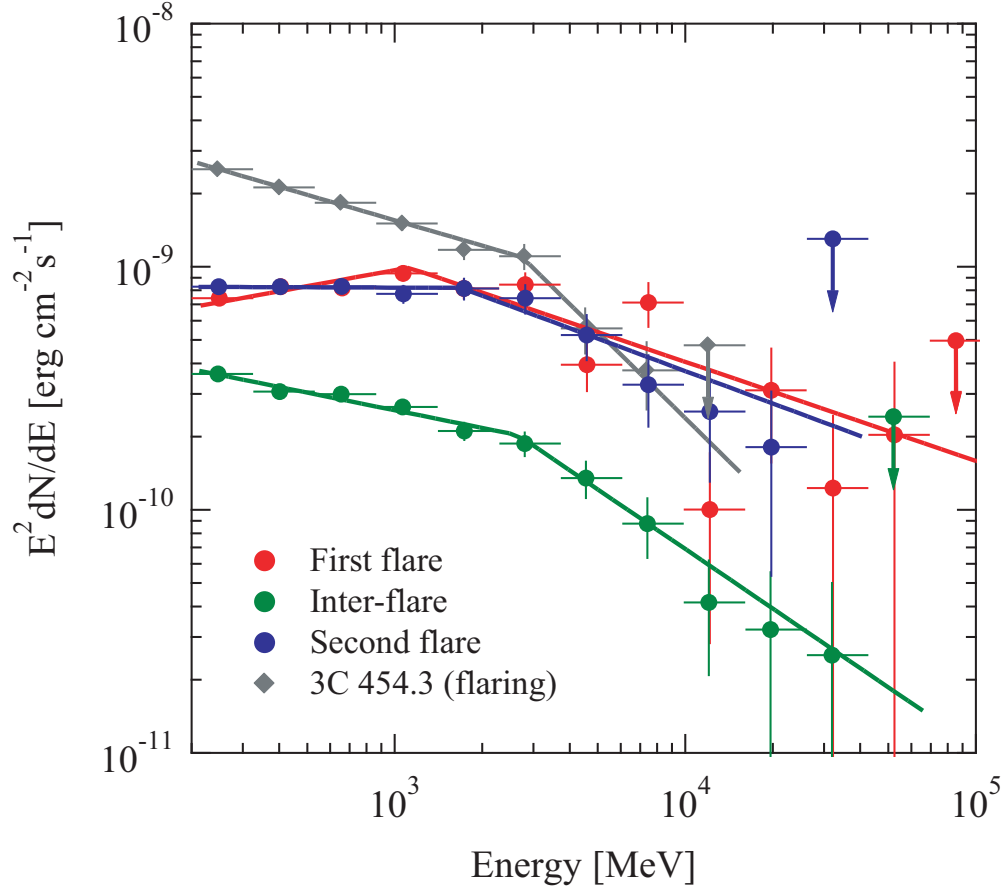}

\caption{The three different spectral energy distributions of 4C\,+21.35 corresponding to different periods of the active state of 4C\,+21.35 (red, blue, and green filled circles), compared with the flaring spectrum of 3C\,454.3 \citep[`the bright December 2009 flare spectrum' of 3C\,454.3 corresponding to the epoch MJD 55166--55173; see][gray diamonds]{Ackermann2010}. Solid lines show the best-fit broken power-law models. Arrows indicate 95\% upper limits.}

\end{figure}

\end{document}